\RequirePackage{fix-cm}
\documentclass[twocolumn]{svjour3}       
\smartqed  
\usepackage{graphicx}
\usepackage{amssymb}
\usepackage{amsmath}
\usepackage[usenames,dvipsnames]{color}
\usepackage{dcolumn}
\usepackage{placeins}
\usepackage{natbib} 
\bibpunct{(}{)}{;}{a}{}{,}

\newcommand{\be}{\begin{equation}}
\newcommand{\ee}{\end{equation}}
\newcommand{\ba}{\begin{eqnarray}}
\newcommand{\ea}{\end{eqnarray}}

\newcommand{\mytilde}{\raise.17ex\hbox{$\scriptstyle\mathtt{\sim}$}}

\journalname{Microfluidics and Nanofluidics}
\begin{document}

\title{Capillary focusing close to a topographic step: Shape and instability of confined liquid filaments
}

\titlerunning{Shape and instability of confined liquid filaments}        

\author{Michael Hein \and
        Shahriar Afkhami \and
        Ralf Seemann \and
        Lou Kondic}

\authorrunning{M. Hein \and S. Afkhami \and R. Seemann \and L. Kondic} 

\institute{M. Hein \at
              Saarland University \\
              Experimental Physics, Saarland University, 66123 Saarbr\"{u}cken, Germany\\
              Tel.: +49-681-302 71704\\
              Fax: +49-681-302 71700\\
              \email{michael.hein@physik.uni-saarland.de}           
           \and
           S. Afkhami \at
              New Jersey Institute of Technology\\
	      Department of Mathematical Sciences, New Jersey Institute of Technology, Newark NJ 07102, USA\\
	      Tel.: +1-973-596-5719\\
              Fax: +1-973-660-6467\\
              \email{shahriar.afkhami@njit.edu}
	   \and
	   R. Seemann \at
              Saarland University \\
	      Experimental Physics, Saarland University, 66123 Saarbr\"{u}cken, Germany\\
              Tel.: +49-681-302 71799\\
              Fax: +49-681-302 71700\\
              \email{r.seemann@physik.uni-saarland.de}
	   \and
           L. Kondic \at
              New Jersey Institute of Technology\\
	      Department of Mathematical Sciences, New Jersey Institute of Technology, Newark NJ 07102, USA\\
	      Tel.: +1-973-596-2996\\
              Fax:  +1-973-660-6467\\
              \email{kondic@njit.edu}		
}

\date{Received: date / Accepted: date}

\maketitle

\begin{abstract}
Step-emulsification is a microfluidic technique for droplet generation which relies on the abrupt decrease of confinement
of a liquid filament surrounded by a continuous phase. A striking feature of this geometry is the transition between two
distinct droplet breakup regimes, the ``step-regime'' and  ``jet-regime'', at a critical capillary number. In the step-regime, 
small and monodisperse droplets break off from the filament directly at a topographic step, while in the jet-regime a jet
protrudes into the larger channel region and large plug-like droplets are produced. We characterize the breakup behavior
as a function of the filament geometry and the capillary number and present experimental results on the shape and evolution
of the filament for a wide range of capillary numbers in the jet-regime. We compare the experimental results with numerical
simulations. Assumptions based on the smallness of the depth of the microfluidic channel allow to reduce the governing equations
to the Hele-Shaw problem with surface tension. The full nonlinear equations are then solved numerically using a volume-of-fluid
based algorithm. The computational framework also captures the transition between both regimes, offering a deeper understanding
of the underlying breakup mechanism.
\keywords{Drops and Bubbles \and Step-Emulsification \and Capillary Focusing \and Hele-Shaw Flow \and Volume-of-Fluid}
\end{abstract}

\section{Introduction}
\label{intro}
\begin{figure}[t!]
\includegraphics[width=41mm]{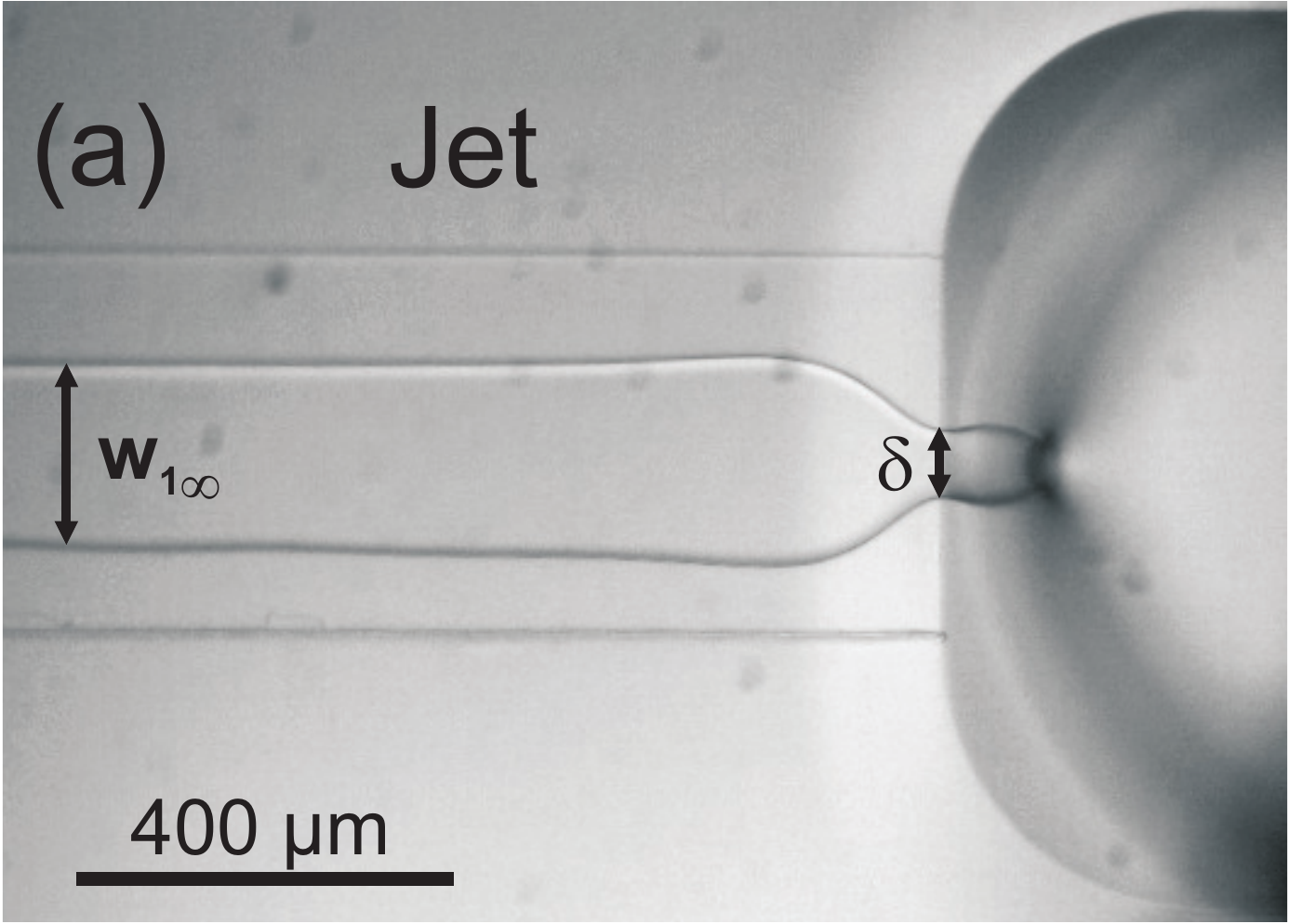}
\includegraphics[width=41mm]{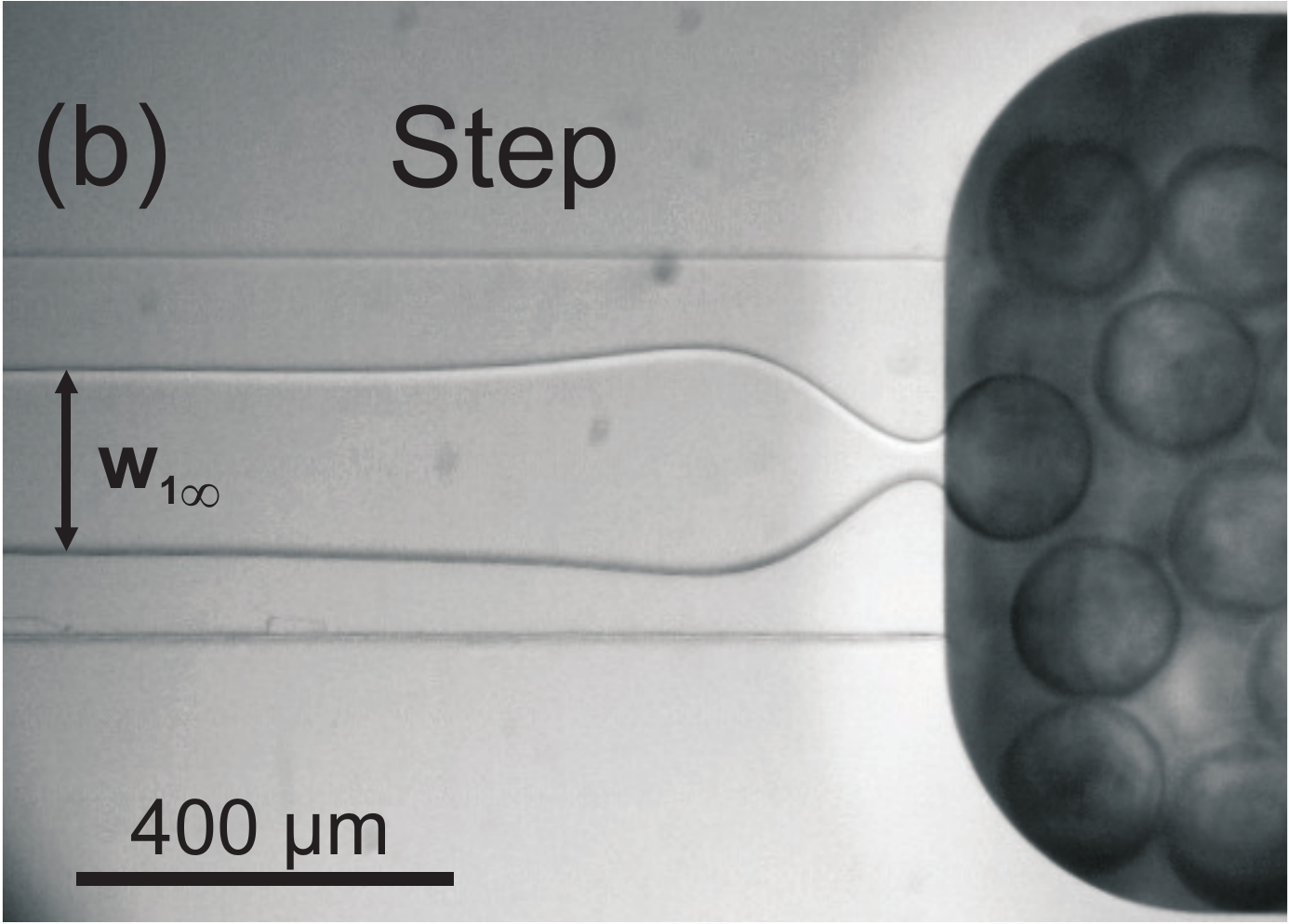}
\caption{Optical micrographs showing two regimes of droplet production.
         (a) ``Jet-emulsification'' for large Ca: Breakup occurs downstream, 
         (b) ``Step-emulsification'' for low Ca: Breakup occurs at the step}
\label{fig:1}
\end{figure}
Hydrodynamic instabilities leading to droplet formation by a decay of a liquid jet or filament have 
received considerable interest over the last two centuries \citep{Rayleigh1878}, renewed especially
since the emergence of two-phase microfluidic systems. Hydrodynamic instabilities 
are employed to generate droplets on the micro-scale, a technique that is promising for many lab-on-a-chip applications, 
where droplets could serve as discrete vials for chemical reactions or bio-analysis. 
Extensive reviews on experimental and theoretical studies of droplet production units and droplet handling in microfluidics 
have been written by \citet{SeemannRev12} and \citet{Woerner2012}. 
Typically three different droplet production units are being used: In a T-Junction \citep{Quake2001}, in which the dispersed phase enters 
a channel filled with a continuous phase from a side channel, droplets are typically produced either by shearing or by a 
``Plug-and-Squeeze mechanism'' \citep{Garstecki2006}. In co-flow or cross-flow geometries, which have received considerable interest from an 
experimental as well as from analytical perspective \citep{Anna2003,Guillot1,Guillot2}, the dispersed phase flows parallel to the continuous 
phase and decays into droplets either directly at the inlet (``Dripping''), forms a jet that decays some distance downstream (``Jetting'') 
or forms a liquid jet, which is absolutely stable (``Co-Flow''). All of these methods are widely used.

The instability of a liquid filament confined in a quasi two-dimensional geometry triggered by a sudden expansion of
the channel has recently been described \citep{Priest2006,Malloggi2010,Humphry2009,Shui2011,Dangla2013}. 
The system consists of a shallow terrace and a larger microfluidic reservoir downstream, as shown in 
Fig.~\ref{fig:1}. In the terrace, the non-wetting dispersed liquid flows as a straight filament confined
at the top and the bottom by the walls and surrounded by a continuous phase. The main control parameter
is the capillary number, Ca, which represents the effect of viscous relative to capillary forces.
Two distinct droplet breakup mechanisms can be observed for droplet generation. At high Ca,
a mechanism called ``jet-emulsification'' occurs, in which the dispersed phase creates a stable tongue
in the terrace. This tongue narrows to a neck close to the topographic step due to capillary focusing. 
From this neck, a jet protrudes, generating large and polydisperse droplets that fill the whole reservoir
channel (see Fig.~\ref{fig:1}a). Below a critical Ca, the so called ``step-emulsification'' occurs (see Fig.~\ref{fig:1}b).
In the latter case, the inner stream forms a droplet immediately after it reaches the topographic step, 
where the tip then becomes unstable, producing droplets at high throughput with a monodispersity superior
to the previously mentioned droplet production methods \citep{Priest2006,Dangla2013}. 
Recent efforts to describe this emulsification process have resulted in models for predicting lower bounds
for the generated droplet size in the step-regime \citep{Dangla2013} and estimating the width of the tongue
at the step in the jet-regime \citep{Malloggi2010}. However, rigorous study of the shape and instability of
the confined tongue and comprehensive comparison of experimental results with theoretical models 
have not been carried out until now.

In this article, we study the capillary focusing phenomena in a step-emulsification microfluidic device. On one hand, the dynamic
filament shape and stability in the jet-regime is experimentally analyzed for a wide parameter range. 
On the other hand, a computational framework is developed to solve the full Hele-Shaw (H-S) equations and to directly compute 
the shape of the liquid-liquid interface in the terrace, providing a better understanding of the observed experiments. 
By direct comparison of the numerical results with experiments,
we show that the shape of the liquid filament in the jet-regime can be computed as a function of Ca and the flow rate ratio between
the dispersed and continuous phase, without resorting to any undetermined parameter. 

Here, a simple stability-criterion is used to show that the experimentally observed transition between the step- and jet-regime
can also be predicted by the numerical simulations; the description of this transition based on the direct numerical solution of the 
two-phase H-S equations with surface tension has not been addressed before.

\section{Experimental method}
\label{exp}
To characterize the droplet breakup regimes in a step-emulsification geometry and the filament shape in the jet-regime, 
experiments are performed using a device micro\-machined into a polymethylmethacrylate (PMMA)-block 
with a terrace of width $w=400$\,$\mu$m and depth $b=30$\,$\mu$m (i.e.~with aspect ratio $w/b=12.76$) and a downstream reservoir of cross-sectional
area of about $1\times1$\,mm$^2$. The channel is sealed by a PMMA-sheet. An aqueous phase with $28\,$wt\% glycerol is injected as
a dispersed phase into an oily continuous phase, IsoparM (ExxonMobil Chemical). For the chosen liquid system, the viscosity of the dispersed phase, $\mu_1$, 
is matched with the viscosity of the continuous phase, $\mu_2$, and is approximately $2.1\,$mPas.
$2\,$wt\% of the surfactant Span $80$ (Sigma Aldrich) are added to the continuous phase to increase the droplet stability and to ensure 
the wetting of the channel walls by the continuous phase. The interfacial tension coefficient, $\gamma$, determined using the pendant drop method,
is $3.5\pm0.1\,$mN/m. The evolution of the interface is imaged using a microscope (Zeiss AxioVert) and a high-speed CMOS camera (PCO 1200hs).
The interface profiles as well as the tongue width, $w_{1\infty}$, far away from the neck, and the neck width, $\delta$, directly at the 
step, are automatically measured using numerical image analysis (see Fig.~\ref{fig:1}a). Volumetric flow rates, $Q_1$ and $Q_2$, of 
the dispersed and the continuous phase, respectively, are adjusted using computer controlled syringe pumps. The corresponding 
$\mbox{Ca} = U_1 \mu_1/\gamma$ is determined from the average flow velocity of the dispersed phase, $U_1$. 

\section{Computational framework}
\label{numerics}
In addition to physical experiments, a computational framework is developed to  
provide a deeper insight into the effects governing the evolution of the liquid-liquid interface in the jet-regime and to predict
the transition between the jet- and step-regime. 
The original governing equations are reduced to the time-dependent two-phase Hele-Shaw equations with surface tension.
We stress that, even in this case, the full problem is nonlinear due to the curvature term in the surface tension
and therefore the full problem has to be solved numerically. Furthermore, although not considered here, 
our numerical model allows to seamlessly vary the viscosity of both phases as well as 
the initial condition and the geometry of the Hele-Shaw problem. We use a volume-of-fluid (VOF) based method  
to directly solve the governing equations. Our numerical model, described in detail by \citet{Afkhami2013},
has the distinctive feature of being capable of, accurately and robustly, modeling the surface tension force.
In addition, the accurate interface reconstruction, the second-order curvature
computation, and the use of adaptive mesh refinement allow for resolved description of the interface, enabling us to 
investigate the complex features of interface profiles extracted from experiments, which was not possible in prior studies.
The order of accuracy and convergence properties of the numerical methods were previously studied by \citet{Afkhami2013}.
\begin{figure}[t!]
  \includegraphics[width=85mm]{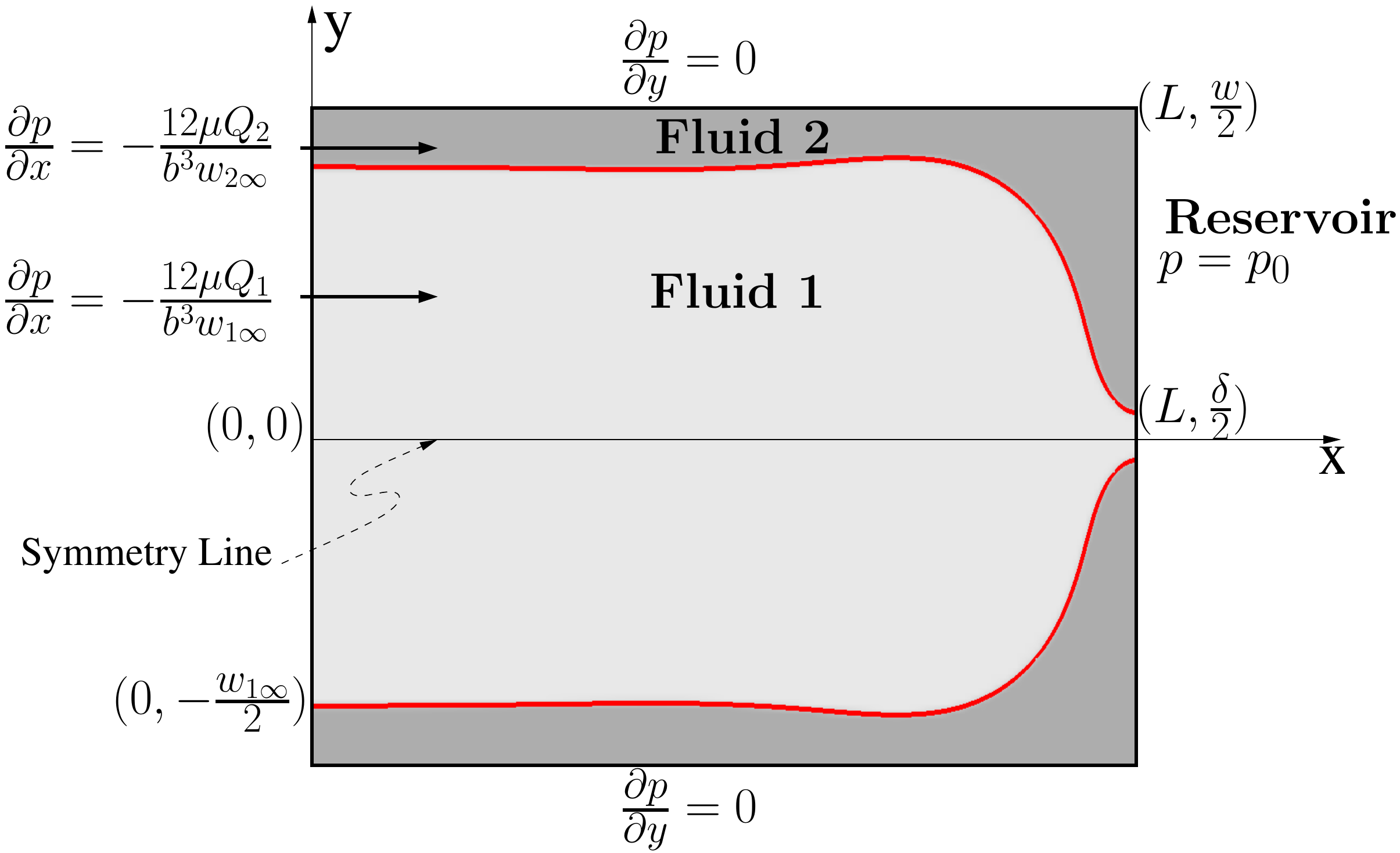}
\caption{
         Schematic of the flow domain for the Hele-Shaw model and
         the corresponding boundary conditions. The domain is bounded
         by walls at  $|y| = w/2$. At $x=0$, fluid $1$
         occupies $|y| \le w_{1\infty}/2$ and fluid $2$ occupies  
         $w_{1\infty}/2 < y \le w$ and  $-w < y <-w_{1\infty}/2$;
         $w_{2\infty} = w-w_{1\infty}$
        }
\label{fig:2}
\end{figure}

The computational framework consists of the classical H-S model to simulate the shape of the interface
between the dispersed and the continuous phase in the shallow channel. 
The depth-averaged velocity field in both phases is defined as
\begin{equation}
{\ensuremath{\mathbf{u}(x,y,t)}} = \frac{b^2}{12 \mu} \left\{-\nabla p(x,y,t) + {\ensuremath{\mathbf{F}}}(x,y,t)\right\},
\label{eq:HS}
\end{equation}
where $\mu$ is the viscosity of the considered phase defined as
\begin{equation}
\ensuremath{\mu(f) = \frac{\mu_1\mu_2}{(1-f)\mu_1 +f\mu_2}},
\label{eq:viscosity}
\end{equation}
where $\mu_1$ and $\mu_2$ refer to the viscosity of fluid $1$ and $2$, respectively,
$b$ is the depth of the H-S cell, and $p(x,y,t)$ is the local pressure.
The VOF function, $f(x,y,t)$, tracks the motion of the interface.
In a VOF method, the discrete form of the function $f$ represents the volume fraction
of a cell filled with, in this case, fluid $1$. Away from the interface, 
$f=0$ (inside fluid $2$) or $f=1$ (inside fluid $1$); ``interface cells'' correspond to $0<f<1$.
The evolution of $f$ satisfies the advection equation 
\begin{equation}
\frac{\partial f}{\partial t} +  {\ensuremath{\mathbf{u}(x,y,t)}} \cdot \nabla f = 0.
\label{eq:f}
\end{equation} 
In this formulation, surface tension enters as a singular body force, 
${\ensuremath{\mathbf{F}}}(x,y,t)$, centered at the interface between two fluids \citep{Afkhami2013}.
Figure \ref{fig:2} is a schematic of the flow domain,
$0\le x\le L$, $|y|\le w/2$, representing the terrace. Due to symmetry, only half of the domain
is simulated. We consider a constant pressure,
$p_0$, in the reservoir and, as a first approximation for the outflow into a reservoir, 
an outflow pressure boundary condition at the exit, $x=L$, as
\begin{equation} 
p(x,y,t) = \left\{
  \begin{array}{l l}
    p_0 + 2A\gamma/b & \quad \mbox{$0<f\le 1$}\\
    p_0 & \quad \mbox{$f=0$}
\end{array} \right.
\label{eq:PressureBC}
\end{equation}
where the out-of-plane curvature is given by $2/b$. Thus $2\gamma/b$ is the Laplace 
pressure inside the dispersed phase $1$. Since the in-plane curvature at the step is typically 
much smaller than the out-of-plane curvature, it can be safely neglected.
A pressure correction parameter, $A \le 1$, is included for modeling the flow configurations 
for which the H-S approximation is expected to be less accurate, as discussed further below.
The boundary condition for $f$ at the top and bottom walls is $f=0$. 
At the outflow, the boundary condition for $f$ is the interface being perpendicular to the 
topographic step: $\partial f/\partial x=0$.
This is justified experimentally in section \ref{evo}. 
On the solid boundaries (side walls), 
$\partial p/\partial y=0$. 
At the inlet, $x=0$, the parallel flow solution holds, meaning that for both
phases, the pressure gradient is prescribed as
$
\partial p(x,y,t)_i/\partial x = -
(12 \mu Q_i)/(b^3 w_{i\infty})
$
where $i=(1,2)$, $w_{i\infty}$ is the width occupied by the $i$th liquid at the inlet,
and $Q_i$ is the flow rate of the $i$th phase.
As in the experiments, we chose $dp_1/dx|_{x=0}=dp_2/dx|_{x=0}$ and fix $b$ and $w$. 
The initial condition at $t=0$ is a flat interface defined by
the initial distribution of $f$ over the domain.
We keep $L$ large enough so the results are unaffected by its value. 

\section{Results and discussion}
\label{results}
\subsection{Filament shape: temporal evolution}
\label{evo}
In the jet-regime, i.e.~at sufficiently high Ca (see Fig.~\ref{fig:1}a), a filament of a dispersed phase, sandwiched by the continuous phase,
is stable over the terrace due to the geometric confinement of the flow. When pushed into the reservoir, in which the pressure is presumably
constant, the two phases are forced to balance their pressure. The dispersed
phase, however, experiences a higher pressure due to the Laplace pressure, generated by the out-of-plane 
curvature of the interface and the evolving meniscus. This induces a temporarily varying cross-stream
pressure difference between the two fluids. Thus, to adjust to a constant
pressure at the step, the dispersed phase has to increase its velocity, whereas the continuous phase moves
more slowly over the terrace. Due to mass conservation, the interface between the two fluids must assume a
tongue-like shape that narrows down to a thin neck along the mean flow, an effect known as ``capillary focusing''.

\begin{figure}[t!]
  \includegraphics[width=95mm,trim=10mm 0 0 0,clip=true]{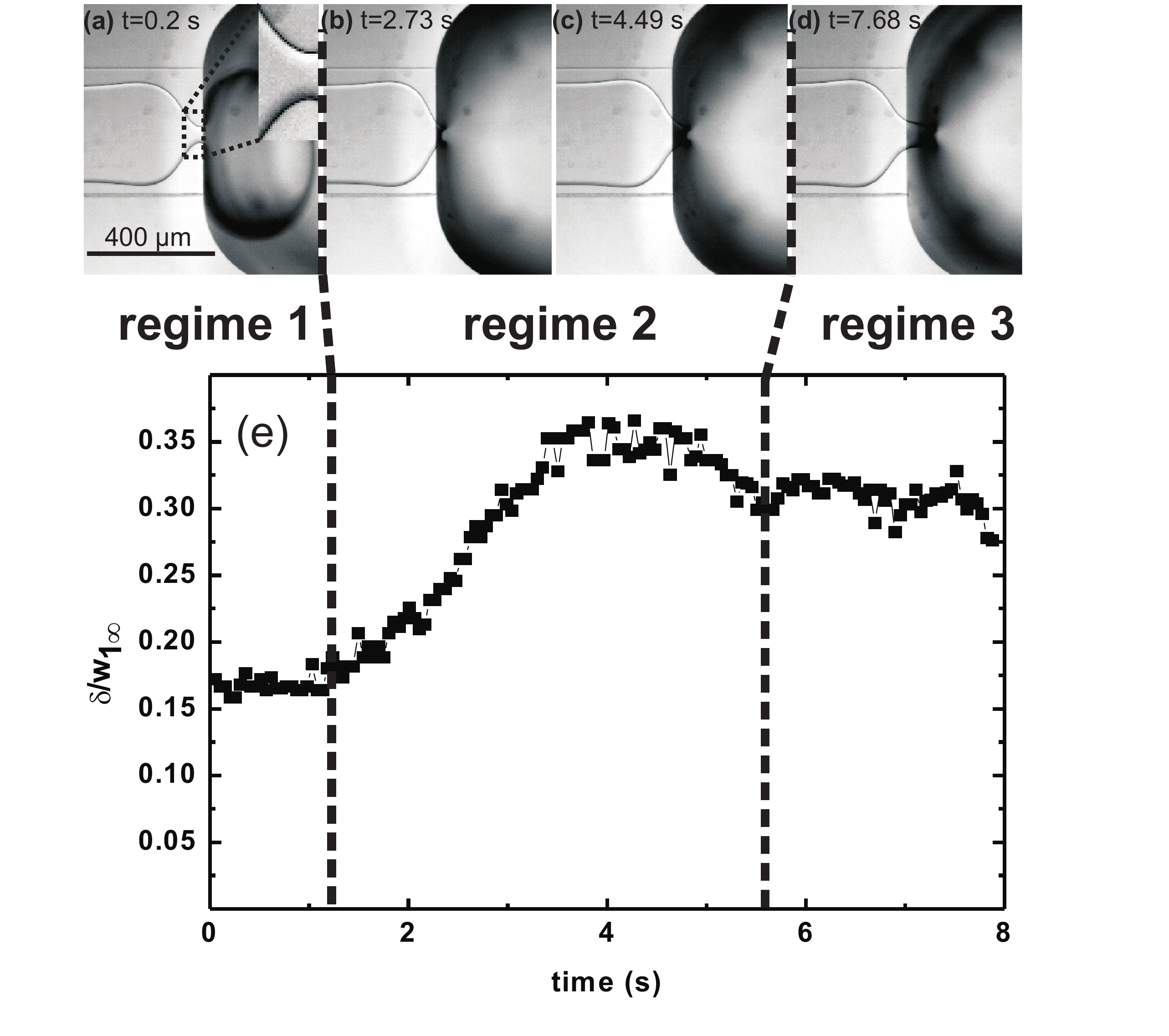}
\caption{Temporal evolution of the tongue in the jet-regime.
         (a)-(d) Micrographs with the inset showing the interface directly at the step and (e) the shape factor, $\delta/w_{1\infty}$, as a function of time; 
         Ca $=9.73\times10^{-3}$ and $Q_{1\infty}/Q_{2\infty}=3$}
\label{fig:3}
\end{figure}
For droplet formation in an actual microfluidic device, the situation is more complex, since droplets are periodically
formed in the reservoir, thus influencing the pressure at the step and consequently the tongue shape. Figure \ref{fig:3}a-d
shows a typical evolution of the tongue shape during droplet production in the jet-regime
(see also the movie in the supplemental material). Figure \ref{fig:3}e shows the tongue width (normalized by
the filament width), $\delta/w_{1\infty}$, called the tongue shape factor, as a function of time. As shown,  
at the beginning of the droplet formation cycle (see Fig.~\ref{fig:3}a and regime 1 in Fig.~\ref{fig:3}e), 
when the droplet is initially pushed into the reservoir, $\delta/w_{1\infty}$ remains constant while the droplet inflates,
maintaining a nearly spherical shape. During this inflation process, the liquid-liquid interface at the step is perpendicular to the 
topographic step (see the inset in Fig.~\ref{fig:3}a). The Laplace pressure that counteracts the inflation of the droplet rapidly 
decreases as the droplet radius increases and remains merely constant when the droplet reaches the reservoir dimension. 
Additional complexity may arise in the case of droplet collision in the reservoir channel. When the droplet touches the walls of the reservoir,
$\delta/w_{1\infty}$ typically increases as the droplet starts to travel downstream, pulling the tongue towards the step 
(see Fig.~\ref{fig:3}b-c and regime 2 in Fig.~\ref{fig:3}e). 
Additionally, the droplet blocks the reservoir, except for small regions along the 
corners of the rectangular reservoir.
Thus, an additional pressure contribution, that is expected to increase with the droplet length \citep{Labrot2009},
is needed to push the continuous phase  at constant flow rate downstream in the reservoir.
This pressure component is also expected to deform the interface in the shallow terrace. 
Thus $\delta/w_{1\infty}$ reaches a maximum during regime 2 in Fig.~\ref{fig:3}e and then decreases,
showing the formation of a neck, that may evolve to a stable width in regime 3 in Fig.~\ref{fig:3}e.
In regime 3, the neck elongates as the rear interface of the droplet travels downstream, until it bulges
and finally ruptures (see Fig.~\ref{fig:3}d and regime 3 in Fig.~\ref{fig:3}e).
After droplet breakup, the process repeats starting in regime 1.
From the above description of the tongue evolution in the jet-regime at high Ca, it is obvious that only in
regime 1 the pressure boundary condition at the topographic step is known a priori, resembling an open outlet
with a constant pressure, $p_0$. Thus the experimental data taken in regime 1 is used for further comparison with computational results.
In regime 2 and 3, however, the pressure at the topographic step is significantly
influenced by the evolving droplet in the reservoir. 
We note that regime 1 can be further expanded by increasing reservoir dimensions. 

\begin{figure}[t!]
\includegraphics[width=87mm]{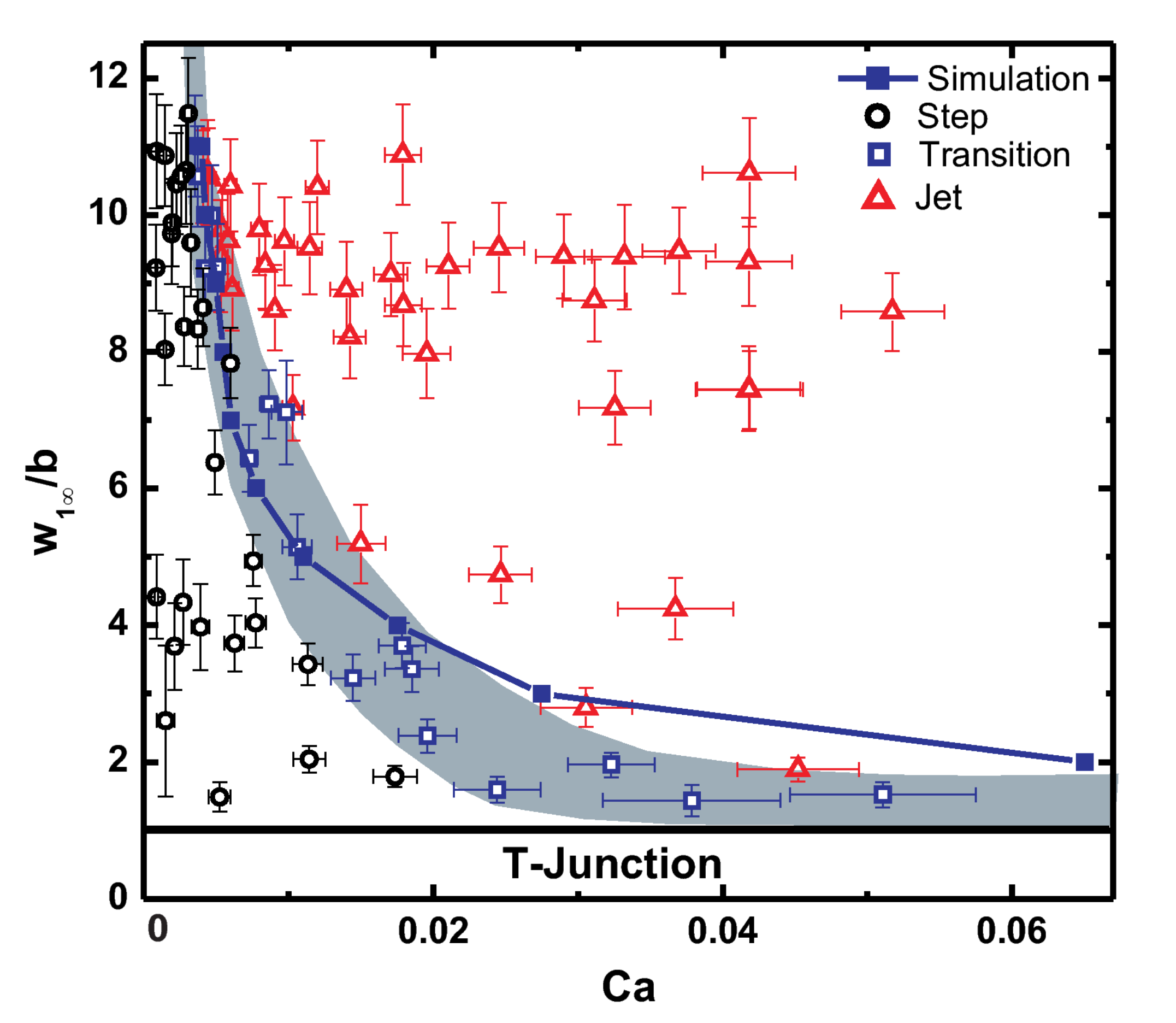}
\caption{Phase diagram demonstrating the transition of jet-regime (red triangles) to step-regime (black circles). 
         Blue open squares denote the transition regime, where droplet production alternates between step- and jet-regime. 
         Blue solid squares indicate the simulation results for $A=0.4$, as discussed in the text.  
         For $w_{1\infty}/b\leq1$ no stable filament is formed
         }
\label{fig:4}
\end{figure}

\subsection{Transition from jet- to step-regime}
\label{trans}
A sudden transition to the step-regime occurs when reducing the Ca below a certain threshold, as shown in Fig.~\ref{fig:4}.
The transition between jet- and step-regime can be qualitatively described as follows:
At low Ca, when the width of the neck, $\delta$, becomes comparable to the height of the terrace, $b$, 
the dispersed phase forms a nearly cylindrical neck, which is prone to a surface-tension-driven instability. 
As the neck is no longer stable, the filament breaks up rapidly and small droplets are continuously produced,
accompanied by a periodic retraction of the tongue. 
We have confirmed this critical neck width experimentally by measuring $\delta$ in the metastable transition regime, 
where step- and jet-regime alternate (see shaded region in Fig.~\ref{fig:4}). 
Interestingly, despite the qualitative differences in tongue shape and stability
between step- and jet-regime (cf. Fig.~\ref{fig:1}), we note that the tongue width far upstream from the step, $w_{1\infty}$, 
stays constant for a given $Q_{1\infty}/Q_{2\infty}$ when varying Ca. In particular, 
$w_{1\infty}$ is unaffected by the transition from the jet- to step-regime, which allows to characterize
the transition as a function of the rescaled tongue width, $w_{1\infty}/b$, and Ca. 
Figure~\ref{fig:4} shows the emulsification phase diagram in $\left( w_{1\infty}/b, \mbox{Ca}\right)$ space, 
with an increased data range compared to~\citet{Priest2006}, in which it is shown that the phase diagram 
is insensitive to the viscosity ratio of the dispersed phase to the continuous phase, $\mu_1/\mu_2$. 
We note that for $w_{1\infty}/b\leq1$, droplet breakup occurs at the inlet of the dispersed phase. 
This regime is comparable to breakup in a T-junction and is not considered here.

\begin{figure}[t!]
\includegraphics[width=88mm]{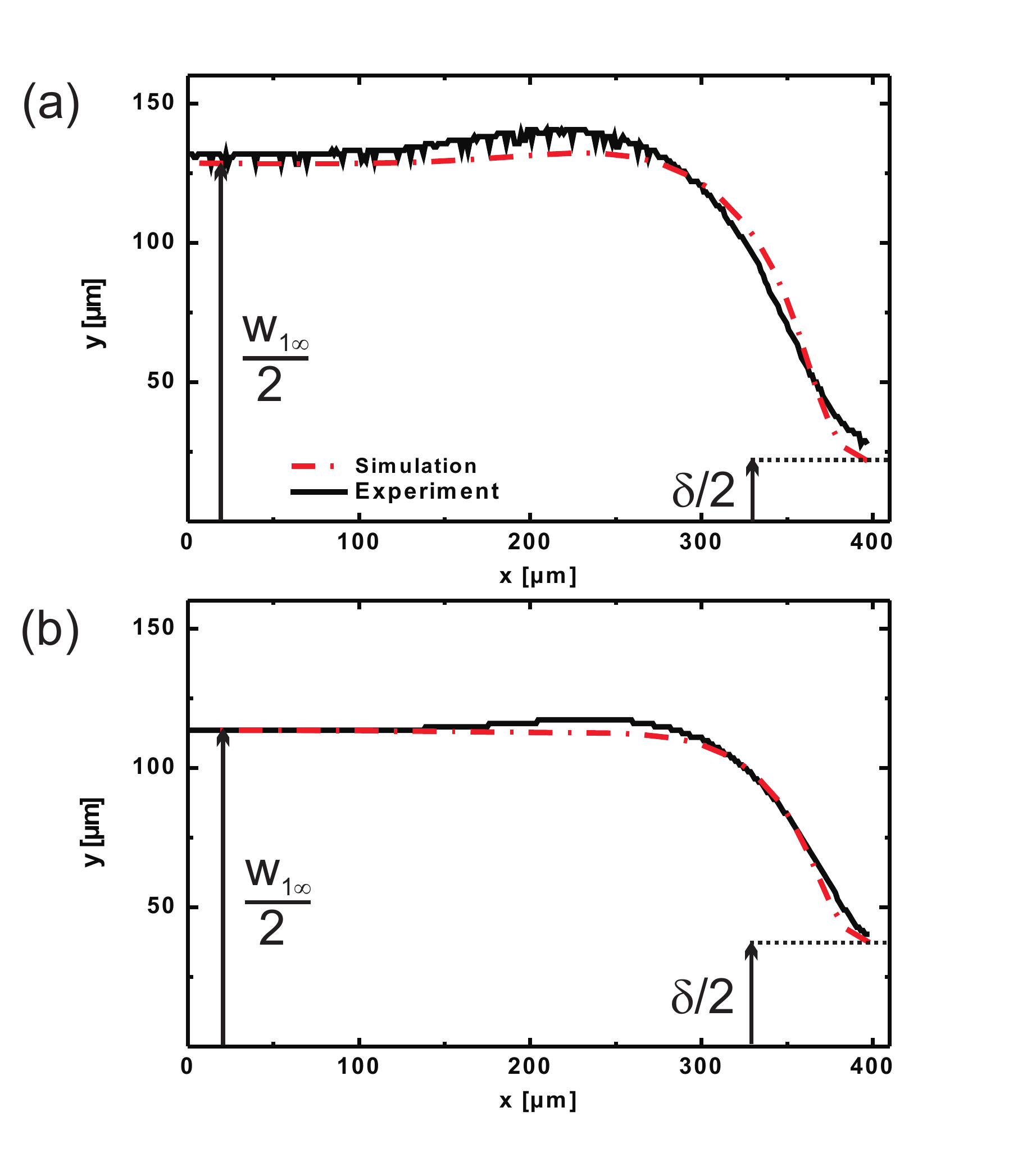}
\caption{Comparison of the tongue profiles in the terrace obtained from experiments (black solid line) 
         and simulations with $A=1$ (red dashed line). Both experimental profiles are obtained in the jet-regime.
         (a) Ca $= 0.019$ and $w_{1\infty}/b = 7.97$  and (b) Ca $= 0.042$ and $w_{1\infty}/b = 7.44$.
         For experiments, $Q_{1\infty}/Q_{2\infty} = 1.5$. For simulations, $Q_{1\infty}/Q_{2\infty}=1.67$ in (a) and 
         $Q_{1\infty}/Q_{2\infty}=1.41$ in (b)}         
\label{fig:5}
\end{figure}
\begin{figure}
\includegraphics[width=90mm]{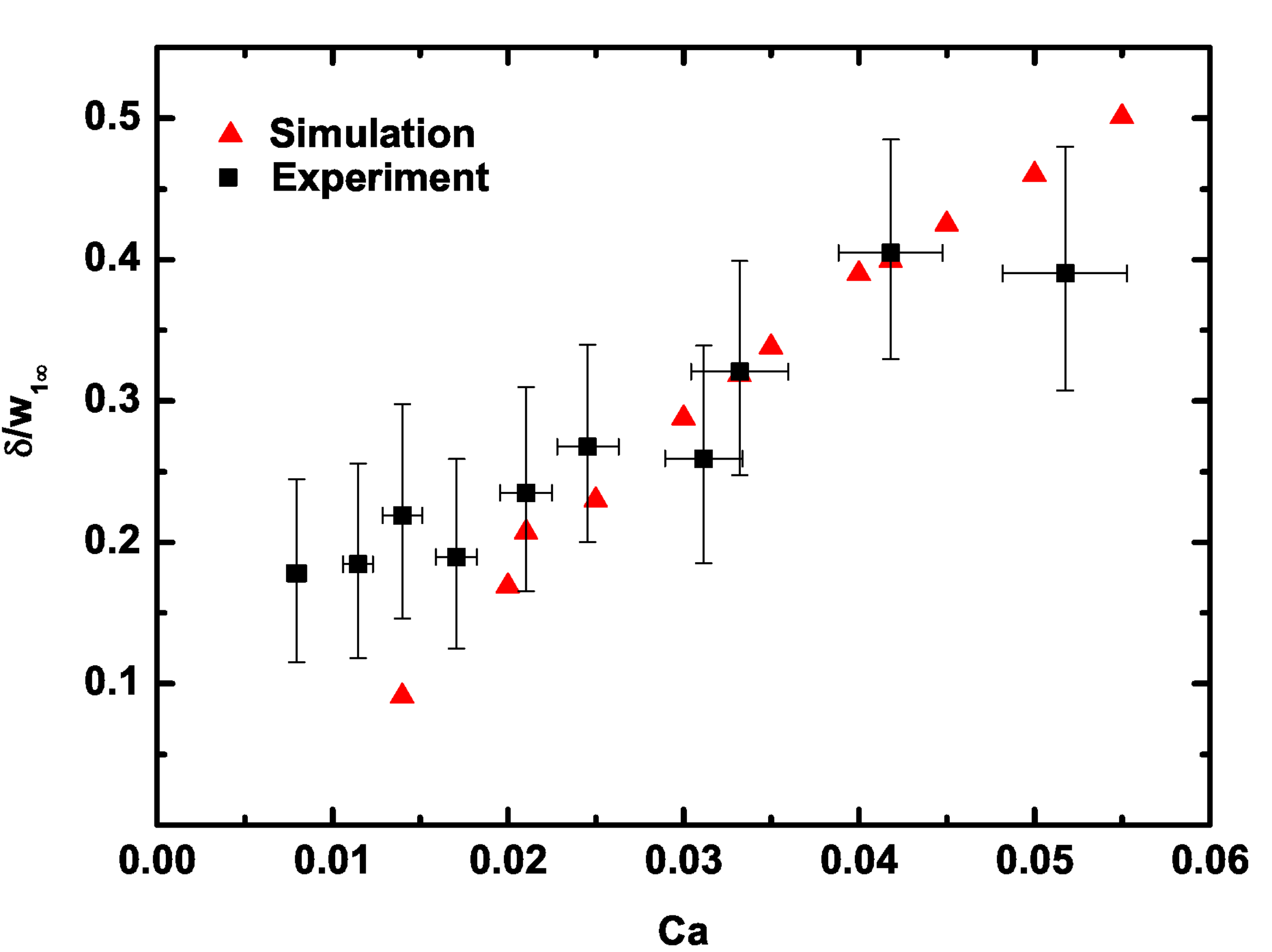}
\caption{Comparison of the experimentally determined shape factor, $\delta/w_{1\infty}$, 
         for $Q_{1\infty}/Q_{2\infty}=3$ (black squares) and the simulation results (red triangles) for $Q_{1\infty}/Q_{2\infty}=2.81$ and $A=1$
	 }
\label{fig:6}
\end{figure}

\subsection{Comparison of experiments with numerical simulations}
\label{comp}
Figure~\ref{fig:5} shows the direct comparison of the tongue profiles obtained from experiments with the corresponding simulations. 
As shown, the numerical simulations accurately reproduce the main features of the tongue shape, particularly the slight increase of
the tongue width followed by the strong capillary focusing close to the topographic step. 
We also note that the agreement between the simulated results and experiments is best at higher Ca values, shown in Fig.~\ref{fig:5}b,
remote from the transition to the step-regime, where the tongue width is larger and the validity of the H-S approximation 
(assumed in the numerical scheme) is clearly fulfilled. To quantify the comparison between the numerical and experimental results,
in Fig.~\ref{fig:6}, the shape factor, $\delta/w_{1\infty}$, measured from the obtained profiles, is plotted as a function of Ca. 
The error bars for the experimentally determined $\delta/w_{1\infty}$ are worst case estimates based on the width of the interface after image 
segmentation, and the error bars for Ca are Gaussian error estimates. The measured values for $\delta/w_{1\infty}$ agree within 
experimental accuracy with the results from the corresponding simulations, especially for medium to high Ca. Both the results 
from the experiment and the simulation show that the neck thickness increases with Ca, as also predicted analytically by \citet{Malloggi2010}.
However, for Ca $\lesssim 0.02$, close to the transition, simulation predicts a stronger focusing, i.e.~a stronger decrease of $\delta/w_{1\infty}$ with Ca, when compared 
to the experimental data, as the neck becomes more and more cylindrical in this regime, making the H-S approximation less accurate. Thus the numerical model 
underestimates $\delta/w_{1\infty}$ for small Ca. This underestimation can be compensated by introducing a pressure correction, $A$, as 
described in Eq.~\ref{eq:PressureBC}, which increases $\delta$ by weakening the capillary focusing, at the cost of a 
reduced bulging prior to the hydrodynamic focusing. 
We apply this pressure correction and use the stability criterion, $\delta=b$, as described in section \ref{trans}, 
and employ the numerical simulation to predict the transition from 
jet- to step-regime. Figure \ref{fig:4} shows the excellent agreement of the computationally predicted transition threshold
when compared to experimental data for $A=0.4$. As discussed previously, introducing $A$ is needed when considering the regime
where the H-S approximation is not strictly valid. A clear approach to improve the accuracy is to consider fully three-dimensional flow, as
well as to consider possible influence of the viscosity contrast and capillary effects on the pressure correction factor.   
We leave these non-trivial extensions for future work.

\section{Conclusion}
\label{conc}
In this article, the capillary focusing of a confined liquid filament at a topographic step is studied both experimentally and numerically.
A novel computational framework, based on the Hele-Shaw approximation and the volume-of-fluid approach, is shown to predict experimental results quantitatively. 
We find that by modeling the reservoir with an applied pressure boundary condition, 
the computed shapes of the filament are in excellent agreement with experiments for moderate to high capillary numbers, without resorting to any undetermined parameters.
Furthermore, we show that the numerical model can accurately predict the width of the tongue remote from the step.
We also characterize the width of the tongue at the step as a function of Ca and show that the 
computational results are in very good agreement with the experimental findings in the jet-regime and close to the transition to the step-regime.
Direct computations also provide an accurate estimate of the transition between two distinct droplet breakup mechanisms characterized experimentally. 
This work is expected to set ground for further numerical and analytical treatment of confined filament shapes in similar geometries. 
The presented computational framework solves the full Hele-Shaw 
equations without additional simplifications. Thus the model is not limited to predicting two-phase flow behavior in the specific geometry discussed in this paper and can
easily be adapted to similar problems. 
Additionally, knowing the dependence of the breakup behaviour on filament geometry and capillary number will also facilitate application-specific 
designs of microfluidic droplet production units. 
\begin{acknowledgements}
Authors gratefully acknowledge Dr. Jean-Baptiste Fleury (Saarland University) for helpful discussions and the DFG-GRK1276 for financial support.
This work was partially supported by the NSF Grant Nos.  DMS-1320037 (S.A.) and CBET-1235710 (L.K.).
\end{acknowledgements}



\bibliography{References}   

\end{document}